\documentstyle{elsart}

\def\fileversion{v1.20a}% was \def\fileversion{v1.20}%
\def\filedate{21.6.94}%  was \def\filedate{26.1.94}%
\edef\epsfigRestoreAt{\catcode`@=\number\catcode`@\relax}%
\catcode`\@=11\relax
\ifx\undefined\@makeother                % -pks-
\def\@makeother#1{\catcode`#1=12\relax}  % -pks-
\fi                                      % -pks-
\immediate\write16{Document style option `epsfig', \fileversion\space
<\filedate> (edited by SPQR + pks)}% was <\filedate> (edited by SPQR)}%
\newcount\EPS@Height \newcount\EPS@Width \newcount\EPS@xscale
\newcount\EPS@yscale
\def\psfigdriver#1{%
  \bgroup\edef\next{\def\noexpand\tempa{#1}}%
    \uppercase\expandafter{\next}%
    \def\LN{DVITOLN03}%
    \def\DVItoPS{DVITOPS}%
    \def\DVIPS{DVIPS}%
    \def\emTeX{EMTEX}%
    \def\OzTeX{OZTEX}%
    \def\Textures{TEXTURES}%
    \global\chardef\fig@driver=0
    \ifx\tempa\LN
        \global\chardef\fig@driver=0\fi
    \ifx\tempa\DVItoPS
        \global\chardef\fig@driver=1\fi
    \ifx\tempa\DVIPS
        \global\chardef\fig@driver=2\fi
    \ifx\tempa\emTeX
        \global\chardef\fig@driver=3\fi
    \ifx\tempa\OzTeX
        \global\chardef\fig@driver=4\fi
    \ifx\tempa\Textures
        \global\chardef\fig@driver=5\fi
  \egroup
\def\psfig@start{}%
\def\psfig@end{}%
\def\epsfig@gofer{}%
\ifcase\fig@driver
\typeout{WARNING! ****
 no specials for LN03 psfig}%
\or % case 1: dvitops
\def\psfig@start{}%
\def\psfig@end{\special{dvitops: import \@p@sfilefinal \space
\@p@swidth sp \space \@p@sheight sp \space fill}%
\if@clip \typeout{Clipping not supported}\fi
\if@angle \typeout{Rotating not supported}\fi
}%
\let\epsfig@gofer\psfig@end
\or %case2 dvips
\def\psfig@start{\special{ps::[begin]  \@p@swidth \space \@p@sheight \space%
        \@p@sbbllx \space \@p@sbblly \space%
        \@p@sbburx \space \@p@sbbury \space%
        startTexFig \space }%
        \if@clip
                \if@verbose
                        \typeout{(clipped to BB) }%
                \fi
                \special{ps:: doclip \space }%
        \fi
        \if@angle              % moved after \if@clip ... \fi -pks-
                \special {ps:: \@p@sangle \space rotate \space}
        \fi
        \special{ps: plotfile \@p@sfilefinal \space }%
        \special{ps::[end] endTexFig \space }%
}%
\def\psfig@end{}%
\def\epsfig@gofer{\if@clip
                        \if@verbose
                           \typeout{(clipped to BB)}%
                        \fi
                        \epsfclipon
                  \fi
                  \epsfsetgraph{\@p@sfilefinal}%
}%
\or % case 3, emTeX
\typeout{WARNING. You must have a .bb info file with the Bounding Box
  of the pcx file}%
\def\psfig@start{}%
\def\psfig@end{\typeout{pcx import of \@p@sfilefinal}%
\if@clip \typeout{Clipping not supported}\fi
\if@angle \typeout{Rotating not supported}\fi
\raisebox{\@p@srheight sp}{\special{em: graph \@p@sfilefinal}}}%
\def\epsfig@gofer{}%
\or % case 4, OzTeX
\def\psfig@start{}%
\def\psfig@end{%
\EPS@Width\@p@swidth
\EPS@Height\@p@sheight
\divide\EPS@Width by 65781  % convert sp to bp
\divide\EPS@Height by 65781
\special{epsf=\@p@sfilefinal
\space
width=\the\EPS@Width
\space
height=\the\EPS@Height
}%
\if@clip \typeout{Clipping not supported}\fi
\if@angle \typeout{Rotating not supported}\fi
}%
\let\epsfig@gofer\psfig@end
\or % case 5, Textures
\def\psfig@end{
         \EPS@Width=\@bbw  
         \divide\EPS@Width by 1000
         \EPS@xscale=\@p@swidth \divide \EPS@xscale by \EPS@Width
         \EPS@Height=\@bbh  
         \divide\EPS@Height by 1000
         \EPS@yscale=\@p@sheight \divide \EPS@yscale by\EPS@Height
  \ifnum\EPS@xscale>\EPS@yscale\EPS@xscale=\EPS@yscale\fi
\if@clip
   \if@verbose
      \typeout{(clipped to BB)}%
   \fi
   \epsfclipon
\fi
\special{illustration \@p@sfilefinal\space scaled \the\EPS@xscale}%
}%
\def\psfig@start{}%
\let\epsfig\psfig
\else
\typeout{WARNING. *** unknown  driver - no psfig}%
\fi
}%
\newdimen\ps@dimcent
\ifx\undefined\fbox
\newdimen\fboxrule
\newdimen\fboxsep
\newdimen\ps@tempdima
\newbox\ps@tempboxa
\long\def\fbox#1{\leavevmode\setbox\ps@tempboxa\hbox{#1}\ps@tempdima\fboxrule
    \advance\ps@tempdima \fboxsep \advance\ps@tempdima \dp\ps@tempboxa
   \hbox{\lower \ps@tempdima\hbox
  {\vbox{\hrule height \fboxrule
          \hbox{\vrule width \fboxrule \hskip\fboxsep
          \vbox{\vskip\fboxsep \box\ps@tempboxa\vskip\fboxsep}\hskip
                 \fboxsep\vrule width \fboxrule}%
                 \hrule height \fboxrule}}}}%
\fi
\ifx\@ifundefined\undefined
\long\def\@ifundefined#1#2#3{\expandafter\ifx\csname
  #1\endcsname\relax#2\else#3\fi}%
\fi
\@ifundefined{typeout}%
{\gdef\typeout#1{\immediate\write\sixt@@n{#1}}}%
{\relax}%
\@ifundefined{epsfig}{}{\typeout{EPSFIG --- already loaded} }%
\@ifundefined{epsfbox}{\input epsf}{}%
\ifx\undefined\@latexerr
        \newlinechar`\^^J
        \def\@spaces{\space\space\space\space}%
        \def\@latexerr#1#2{%
        \edef\@tempc{#2}\expandafter\errhelp\expandafter{\@tempc}%
        \typeout{Error. \space see a manual for explanation.^^J
         \space\@spaces\@spaces\@spaces Type \space H <return> \space for
         immediate help.}\errmessage{#1}}%
\fi
\def\@whattodo{You tried to include a PostScript figure which
cannot be found^^JIf you press return to carry on anyway,^^J
The failed name will be printed in place of the figure.^^J
or type X to quit}%
\def\@whattodobb{You tried to include a PostScript figure which
has no^^Jbounding box, and you supplied none.^^J
If you press return to carry on anyway,^^J
The failed name will be printed in place of the figure.^^J
or type X to quit}%
\def\@nnil{\@nil}%
\def\@empty{}%
\def\@psdonoop#1\@@#2#3{}%
\def\@psdo#1:=#2\do#3{\edef\@psdotmp{#2}\ifx\@psdotmp\@empty \else
    \expandafter\@psdoloop#2,\@nil,\@nil\@@#1{#3}\fi}%
\def\@psdoloop#1,#2,#3\@@#4#5{\def#4{#1}\ifx #4\@nnil \else
       #5\def#4{#2}\ifx #4\@nnil \else#5\@ipsdoloop #3\@@#4{#5}\fi\fi}%
\def\@ipsdoloop#1,#2\@@#3#4{\def#3{#1}\ifx #3\@nnil
       \let\@nextwhile=\@psdonoop \else
      #4\relax\let\@nextwhile=\@ipsdoloop\fi\@nextwhile#2\@@#3{#4}}%
\def\@tpsdo#1:=#2\do#3{\xdef\@psdotmp{#2}\ifx\@psdotmp\@empty \else
    \@tpsdoloop#2\@nil\@nil\@@#1{#3}\fi}%
\def\@tpsdoloop#1#2\@@#3#4{\def#3{#1}\ifx #3\@nnil
       \let\@nextwhile=\@psdonoop \else
      #4\relax\let\@nextwhile=\@tpsdoloop\fi\@nextwhile#2\@@#3{#4}}%
\long\def\epsfaux#1#2:#3\\{\ifx#1\epsfpercent
   \def\testit{#2}\ifx\testit\epsfbblit
        \@atendfalse
        \epsf@atend #3 . \\%
        \if@atend
           \if@verbose
                \typeout{epsfig: found `(atend)'; continuing search}%
           \fi
        \else
                \epsfgrab #3 . . . \\%
                \epsffileokfalse\global\no@bbfalse
                \global\epsfbbfoundtrue
        \fi
   \fi\fi}%
\def\epsf@atendlit{(atend)}
\def\epsf@atend #1 #2 #3\\{%
   \def\epsf@tmp{#1}\ifx\epsf@tmp\empty
      \epsf@atend #2 #3 .\\\else
   \ifx\epsf@tmp\epsf@atendlit\@atendtrue\fi\fi}%

\chardef\trig@letter = 11
\chardef\other = 12
 
\newif\ifdebug %%% turn me on to see TeX hard at work ...
\newif\ifc@mpute %%% don't need to compute some values
\newif\if@atend
\c@mputetrue % but assume that we do
 
\let\then = \relax
\def\r@dian{pt }%
\let\r@dians = \r@dian
\let\dimensionless@nit = \r@dian
\let\dimensionless@nits = \dimensionless@nit
\def\internal@nit{sp }%
\let\internal@nits = \internal@nit
\newif\ifstillc@nverging
\def \Mess@ge #1{\ifdebug \then \message {#1} \fi}%
 
{ %%% Things that need abnormal catcodes %%%
        \catcode `\@ = \trig@letter
        \gdef \nodimen {\expandafter \n@dimen \the \dimen}%
        \gdef \term #1 #2 #3%
               {\edef \t@ {\the #1}%%% freeze parameter 1 (count, by value)
                \edef \t@@ {\expandafter \n@dimen \the #2\r@dian}%
                \t@rm {\t@} {\t@@} {#3}%
               }%
        \gdef \t@rm #1 #2 #3%
               {{%
                \count 0 = 0
                \dimen 0 = 1 \dimensionless@nit
                \dimen 2 = #2\relax
                \Mess@ge {Calculating term #1 of \nodimen 2}%
                \loop
                \ifnum  \count 0 < #1
                \then   \advance \count 0 by 1
                        \Mess@ge {Iteration \the \count 0 \space}%
                        \Multiply \dimen 0 by {\dimen 2}%
                        \Mess@ge {After multiplication, term = \nodimen 0}%
                        \Divide \dimen 0 by {\count 0}%
                        \Mess@ge {After division, term = \nodimen 0}%
                \repeat
                \Mess@ge {Final value for term #1 of
                                \nodimen 2 \space is \nodimen 0}%
                \xdef \Term {#3 = \nodimen 0 \r@dians}%
                \aftergroup \Term
               }}%
        \catcode `\p = \other
        \catcode `\t = \other
        \gdef \n@dimen #1pt{#1} %%% throw away the ``pt''
}%
 
\def \Divide #1by #2{\divide #1 by #2} %%% just a synonym
 
\def \Multiply #1by #2%%% allows division of a dimen by a dimen
       {{%%% should really freeze parameter 2 (dimen, passed by value)
        \count 0 = #1\relax
        \count 2 = #2\relax
        \count 4 = 65536
        \Mess@ge {Before scaling, count 0 = \the \count 0 \space and
                        count 2 = \the \count 2}%
        \ifnum  \count 0 > 32767 %%% do our best to avoid overflow
        \then   \divide \count 0 by 4
                \divide \count 4 by 4
        \else   \ifnum  \count 0 < -32767
                \then   \divide \count 0 by 4
                        \divide \count 4 by 4
                \else
                \fi
        \fi
        \ifnum  \count 2 > 32767 %%% while retaining reasonable accuracy
        \then   \divide \count 2 by 4
                \divide \count 4 by 4
        \else   \ifnum  \count 2 < -32767
                \then   \divide \count 2 by 4
                        \divide \count 4 by 4
                \else
                \fi
        \fi
        \multiply \count 0 by \count 2
        \divide \count 0 by \count 4
        \xdef \product {#1 = \the \count 0 \internal@nits}%
        \aftergroup \product
       }}%
 
\def\r@duce{\ifdim\dimen0 > 90\r@dian \then   % sin(x) = sin(180-x)
                \multiply\dimen0 by -1
                \advance\dimen0 by 180\r@dian
                \r@duce
            \else \ifdim\dimen0 < -90\r@dian \then  % sin(x) = sin(360+x)
                \advance\dimen0 by 360\r@dian
                \r@duce
                \fi
            \fi}%
 
\def\Sine#1%
       {{%
        \dimen 0 = #1 \r@dian
        \r@duce
        \ifdim\dimen0 = -90\r@dian \then
           \dimen4 = -1\r@dian
           \c@mputefalse
        \fi
        \ifdim\dimen0 = 90\r@dian \then
           \dimen4 = 1\r@dian
           \c@mputefalse
        \fi
        \ifdim\dimen0 = 0\r@dian \then
           \dimen4 = 0\r@dian
           \c@mputefalse
        \fi
        \ifc@mpute \then
                \divide\dimen0 by 180
                \dimen0=3.141592654\dimen0
                \dimen 2 = 3.1415926535897963\r@dian %%% a well-known constant
                \divide\dimen 2 by 2 %%% we only deal with -pi/2 : pi/2
                \Mess@ge {Sin: calculating Sin of \nodimen 0}%
                \count 0 = 1 %%% see power-series expansion for sine
                \dimen 2 = 1 \r@dian %%% ditto
                \dimen 4 = 0 \r@dian %%% ditto
                \loop
                        \ifnum  \dimen 2 = 0 %%% then we've done
                        \then   \stillc@nvergingfalse
                        \else   \stillc@nvergingtrue
                        \fi
                        \ifstillc@nverging %%% then calculate next term
                        \then   \term {\count 0} {\dimen 0} {\dimen 2}%
                                \advance \count 0 by 2
                                \count 2 = \count 0
                                \divide \count 2 by 2
                                \ifodd  \count 2 %%% signs alternate
                                \then   \advance \dimen 4 by \dimen 2
                                \else   \advance \dimen 4 by -\dimen 2
                                \fi
                \repeat
        \fi
                        \xdef \sine {\nodimen 4}%
       }}%
 
\def\Cosine#1{\ifx\sine\UnDefined\edef\Savesine{\relax}\else
                             \edef\Savesine{\sine}\fi
        {\dimen0=#1\r@dian\multiply\dimen0 by -1
         \advance\dimen0 by 90\r@dian
         \Sine{\nodimen 0}%
         \xdef\cosine{\sine}%
         \xdef\sine{\Savesine}}}
\def\psdraft{\def\@psdraft{0}}%
\def\psfull{\def\@psdraft{1}}%
\psfull
\newif\if@compress
\def\pscompress{\@compresstrue}
\def\psnocompress{\@compressfalse}
\@compressfalse
\newif\if@scalefirst
\def\psscalefirst{\@scalefirsttrue}%
\def\psrotatefirst{\@scalefirstfalse}%
\psrotatefirst
\newif\if@draftbox
\def\psnodraftbox{\@draftboxfalse}%
\@draftboxtrue
\newif\if@noisy
\@noisyfalse
\newif\ifno@bb
\newif\if@bbllx
\newif\if@bblly
\newif\if@bburx
\newif\if@bbury
\newif\if@height
\newif\if@width
\newif\if@rheight
\newif\if@rwidth
\newif\if@angle
\newif\if@clip
\newif\if@verbose
\newif\if@prologfile
\def\@p@@sprolog#1{\@prologfiletrue\def\@prologfileval{#1}}%
\def\@p@@sclip#1{\@cliptrue}%
\newif\ifepsfig@dos  % only single suffix possible
\def\epsfigdos{\epsfig@dostrue}%
\epsfig@dosfalse
\newif\ifuse@psfig
\def\ParseName#1{\expandafter\@Parse#1}%
\def\@Parse#1.#2:{\gdef\BaseName{#1}\gdef\FileType{#2}}%

\def\@p@@sfile#1{%
  \ifepsfig@dos
     \ParseName{#1:}%
  \else
     \gdef\BaseName{#1}\gdef\FileType{}%
  \fi
  \def\@p@sfile{NO FILE: #1}%
  \def\@p@sfilefinal{NO FILE: #1}%
  \openin1=#1
  \ifeof1\closein1\openin1=\BaseName.bb
    \ifeof1\closein1
      \if@bbllx                 % No postscript file but bb given explicitly.
        \if@bblly\if@bburx\if@bbury
          \def\@p@sfile{#1}%
          \def\@p@sfilefinal{#1}%
        \fi\fi\fi
      \else                     % No bounding box found.
        \@latexerr{ERROR. PostScript file #1 not found}\@whattodo
        \@p@@sbbllx{100bp}%
        \@p@@sbblly{100bp}%
        \@p@@sbburx{200bp}%
        \@p@@sbbury{200bp}%
        \psdraft
      \fi
    \else                       % Postscript file is compressed.
      \closein1%
      \edef\@p@sfile{\BaseName.bb}%
      \typeout{using BB from \@p@sfile}%
      \ifnum\fig@driver=3
        \edef\@p@sfilefinal{\BaseName.pcx}%
      \else
        \ifepsfig@dos
          \edef\@p@sfilefinal{"`gunzip -c `texfind \BaseName.{z,Z,gz}"}%
        \else
          \edef\@p@sfilefinal{"`epsfig \if@compress-c \fi#1"}%          
        \fi
      \fi
    \fi
  \else\closein1                % Postscript file is not compressed.
    \edef\@p@sfile{#1}%
    \if@compress  
      \edef\@p@sfilefinal{"`epsfig -c #1"}%
    \else
      \edef\@p@sfilefinal{#1}%
    \fi
  \fi%
}

\let\@p@@sfigure\@p@@sfile
\def\@p@@sbbllx#1{%
                                            \@bbllxtrue
                \ps@dimcent=#1
                \edef\@p@sbbllx{\number\ps@dimcent}%
                \divide\ps@dimcent by65536
                \global\edef\epsfllx{\number\ps@dimcent}%
}%
\def\@p@@sbblly#1{%
                \@bbllytrue
                \ps@dimcent=#1
                \edef\@p@sbblly{\number\ps@dimcent}%
                \divide\ps@dimcent by65536
                \global\edef\epsflly{\number\ps@dimcent}%
}%
\def\@p@@sbburx#1{%
                \@bburxtrue
                \ps@dimcent=#1
                \edef\@p@sbburx{\number\ps@dimcent}%
                \divide\ps@dimcent by65536
                \global\edef\epsfurx{\number\ps@dimcent}%
}%
\def\@p@@sbbury#1{%
                \@bburytrue
                \ps@dimcent=#1
                \edef\@p@sbbury{\number\ps@dimcent}%
                \divide\ps@dimcent by65536
                \global\edef\epsfury{\number\ps@dimcent}%
}%
\def\@p@@sheight#1{%
                \@heighttrue
                \global\epsfysize=#1
                \ps@dimcent=#1
                \edef\@p@sheight{\number\ps@dimcent}%
}%
\def\@p@@swidth#1{%
                \@widthtrue
                \global\epsfxsize=#1
                \ps@dimcent=#1
                \edef\@p@swidth{\number\ps@dimcent}% 
}%
\def\@p@@srheight#1{%
                \@rheighttrue\use@psfigtrue
                \ps@dimcent=#1
                \edef\@p@srheight{\number\ps@dimcent}%
}%
\def\@p@@srwidth#1{%
                \@rwidthtrue\use@psfigtrue
                \ps@dimcent=#1
                \edef\@p@srwidth{\number\ps@dimcent}%
}%
\def\@p@@sangle#1{%
                \use@psfigtrue
                \@angletrue
                \edef\@p@sangle{#1}%
}%
\def\@p@@ssilent#1{%
                \@verbosefalse
}%
\def\@p@@snoisy#1{%
                \@verbosetrue
}%
\def\@cs@name#1{\csname #1\endcsname}%
\def\@setparms#1=#2,{\@cs@name{@p@@s#1}{#2}}%
\def\ps@init@parms{%
                \@bbllxfalse \@bbllyfalse
                \@bburxfalse \@bburyfalse
                \@heightfalse \@widthfalse
                \@rheightfalse \@rwidthfalse
                \def\@p@sbbllx{}\def\@p@sbblly{}%
                \def\@p@sbburx{}\def\@p@sbbury{}%
                \def\@p@sheight{}\def\@p@swidth{}%
                \def\@p@srheight{}\def\@p@srwidth{}%
                \def\@p@sangle{0}%
                \def\@p@sfile{}%
                \use@psfigfalse
                \@prologfilefalse
                \def\@sc{}%
                \if@noisy
                        \@verbosetrue
                \else
                        \@verbosefalse
                \fi
                \@clipfalse
}%
\def\parse@ps@parms#1{%
                \@psdo\@psfiga:=#1\do
                   {\expandafter\@setparms\@psfiga,}%
\if@prologfile
\fi
}%
\def\bb@missing{%
        \if@verbose
            \typeout{psfig: searching \@p@sfile \space  for bounding box}%
        \fi
        \epsfgetbb{\@p@sfile}%
        \ifepsfbbfound
            \ps@dimcent=\epsfllx bp\edef\@p@sbbllx{\number\ps@dimcent}%
            \ps@dimcent=\epsflly bp\edef\@p@sbblly{\number\ps@dimcent}%
            \ps@dimcent=\epsfurx bp\edef\@p@sbburx{\number\ps@dimcent}%
            \ps@dimcent=\epsfury bp\edef\@p@sbbury{\number\ps@dimcent}%
        \else
            \epsfbbfoundfalse
        \fi
}
\newdimen\p@intvaluex
\newdimen\p@intvaluey
\def\rotate@#1#2{{\dimen0=#1 sp\dimen1=#2 sp
                  \global\p@intvaluex=\cosine\dimen0
                  \dimen3=\sine\dimen1
                  \global\advance\p@intvaluex by -\dimen3
                  \global\p@intvaluey=\sine\dimen0
                  \dimen3=\cosine\dimen1
                  \global\advance\p@intvaluey by \dimen3
                  }}%
\def\compute@bb{%
                \epsfbbfoundfalse
                \if@bbllx\epsfbbfoundtrue\fi
                \if@bblly\epsfbbfoundtrue\fi
                \if@bburx\epsfbbfoundtrue\fi
                \if@bbury\epsfbbfoundtrue\fi
                \ifepsfbbfound\else\bb@missing\fi
                \ifepsfbbfound\else
                \@latexerr{ERROR. cannot locate BoundingBox}\@whattodobb
                        \@p@@sbbllx{100bp}%
                        \@p@@sbblly{100bp}%
                        \@p@@sbburx{200bp}%
                        \@p@@sbbury{200bp}%
                        \no@bbtrue
                        \psdraft
                \fi
                \count203=\@p@sbburx
                \count204=\@p@sbbury
                \advance\count203 by -\@p@sbbllx
                \advance\count204 by -\@p@sbblly
                \edef\ps@bbw{\number\count203}%
                \edef\ps@bbh{\number\count204}%
                 \edef\@bbw{\number\count203}%
                \edef\@bbh{\number\count204}%
               \if@angle
                        \Sine{\@p@sangle}\Cosine{\@p@sangle}%
 
{\ps@dimcent=\maxdimen\xdef\r@p@sbbllx{\number\ps@dimcent}%
 
\xdef\r@p@sbblly{\number\ps@dimcent}%
 
\xdef\r@p@sbburx{-\number\ps@dimcent}%
 
\xdef\r@p@sbbury{-\number\ps@dimcent}}%
                        \def\minmaxtest{%
                           \ifnum\number\p@intvaluex<\r@p@sbbllx
                              \xdef\r@p@sbbllx{\number\p@intvaluex}\fi
                           \ifnum\number\p@intvaluex>\r@p@sbburx
                              \xdef\r@p@sbburx{\number\p@intvaluex}\fi
                           \ifnum\number\p@intvaluey<\r@p@sbblly
                              \xdef\r@p@sbblly{\number\p@intvaluey}\fi
                           \ifnum\number\p@intvaluey>\r@p@sbbury
                              \xdef\r@p@sbbury{\number\p@intvaluey}\fi
                           }%
                        \rotate@{\@p@sbbllx}{\@p@sbblly}%
                        \minmaxtest
                        \rotate@{\@p@sbbllx}{\@p@sbbury}%
                        \minmaxtest
                        \rotate@{\@p@sbburx}{\@p@sbblly}%
                        \minmaxtest
                        \rotate@{\@p@sbburx}{\@p@sbbury}%
                        \minmaxtest
 
\edef\@p@sbbllx{\r@p@sbbllx}\edef\@p@sbblly{\r@p@sbblly}%
 
\edef\@p@sbburx{\r@p@sbburx}\edef\@p@sbbury{\r@p@sbbury}%
                \fi
                \count203=\@p@sbburx
                \count204=\@p@sbbury
                \advance\count203 by -\@p@sbbllx
                \advance\count204 by -\@p@sbblly
                \edef\@bbw{\number\count203}%
                \edef\@bbh{\number\count204}%
}%
\def\in@hundreds#1#2#3{\count240=#2 \count241=#3
                     \count100=\count240        % 100 is first digit #2/#3
                     \divide\count100 by \count241
                     \count101=\count100
                     \multiply\count101 by \count241
                     \advance\count240 by -\count101
                     \multiply\count240 by 10
                     \count101=\count240        %101 is second digit of #2/#3
                     \divide\count101 by \count241
                     \count102=\count101
                     \multiply\count102 by \count241
                     \advance\count240 by -\count102
                     \multiply\count240 by 10
                     \count102=\count240        % 102 is the third digit
                     \divide\count102 by \count241
                     \count200=#1\count205=0
                     \count201=\count200
                        \multiply\count201 by \count100
                        \advance\count205 by \count201
                     \count201=\count200
                        \divide\count201 by 10
                        \multiply\count201 by \count101
                        \advance\count205 by \count201
                     \count201=\count200
                        \divide\count201 by 100
                        \multiply\count201 by \count102
                        \advance\count205 by \count201
                     \edef\@result{\number\count205}%
}%
\def\compute@wfromh{%
                \in@hundreds{\@p@sheight}{\@bbw}{\@bbh}%
                \edef\@p@swidth{\@result}%
}%
\def\compute@hfromw{%
                \in@hundreds{\@p@swidth}{\@bbh}{\@bbw}%
                \edef\@p@sheight{\@result}%
}%
\def\compute@handw{%
                \if@height
                        \if@width
                        \else
                                \compute@wfromh
                        \fi
                \else
                        \if@width
                                \compute@hfromw
                        \else
                                \edef\@p@sheight{\@bbh}%
                                \edef\@p@swidth{\@bbw}%
                        \fi
                \fi
}%
\def\compute@resv{%
                \if@rheight \else \edef\@p@srheight{\@p@sheight} \fi
                \if@rwidth \else \edef\@p@srwidth{\@p@swidth} \fi
}%
\def\compute@sizes{%
        \if@scalefirst\if@angle
        \if@width
           \in@hundreds{\@p@swidth}{\@bbw}{\ps@bbw}%
           \edef\@p@swidth{\@result}%
        \fi
        \if@height
           \in@hundreds{\@p@sheight}{\@bbh}{\ps@bbh}%
           \edef\@p@sheight{\@result}%
        \fi
        \fi\fi
        \compute@handw
        \compute@resv
}
\long\def\graphic@verb#1{\def\next{#1}%
  {\expandafter\graphic@strip\meaning\next}}
\def\graphic@strip#1>{}
\def\graphic@zapspace#1{%
  #1\ifx\graphic@zapspace#1\graphic@zapspace%
  \else\expandafter\graphic@zapspace%
  \fi}
\def\psfig#1{%
\edef\@tempa{\graphic@zapspace#1{}}%
\ifvmode\leavevmode\fi\vbox {%
        \ps@init@parms
        \parse@ps@parms{\@tempa}%
        \ifnum\@psdraft=1
                \typeout{[\@p@sfilefinal]}%
                \if@verbose
                        \typeout{epsfig: using PSFIG macros}%
                \fi
                \psfig@method
        \else
                \epsfig@draft
        \fi
}
}%
\def\graphic@zapspace#1{%
  #1\ifx\graphic@zapspace#1\graphic@zapspace%
  \else\expandafter\graphic@zapspace%
  \fi}
\def\epsfig#1{%
\edef\@tempa{\graphic@zapspace#1{}}%
\ifvmode\leavevmode\fi\vbox {%
        \ps@init@parms
        \parse@ps@parms{\@tempa}%
        \ifnum\@psdraft=1
          \if@angle\use@psfigtrue\fi
          {\ifnum\fig@driver=1\global\use@psfigtrue\fi}%
          {\ifnum\fig@driver=3\global\use@psfigtrue\fi}%
          {\ifnum\fig@driver=4\global\use@psfigtrue\fi}%
          {\ifnum\fig@driver=5\global\use@psfigtrue\fi}%
                \ifuse@psfig
                        \if@verbose
                                \typeout{epsfig: using PSFIG macros}%
                        \fi
                        \psfig@method
                \else
                        \if@verbose
                                \typeout{epsfig: using EPSF macros}%
                        \fi
                        \epsf@method
                \fi
        \else
                \epsfig@draft
        \fi
}%
}%

\def\epsf@method{%
        \epsfbbfoundfalse
        \if@bbllx\epsfbbfoundtrue\fi
        \if@bblly\epsfbbfoundtrue\fi
        \if@bburx\epsfbbfoundtrue\fi
        \if@bbury\epsfbbfoundtrue\fi
        \ifepsfbbfound\else\epsfgetbb{\@p@sfile}\fi
        \ifepsfbbfound
           \typeout{<\@p@sfilefinal>}%
           \epsfig@gofer
        \else
          \@latexerr{ERROR - Cannot locate BoundingBox}\@whattodobb
          \@p@@sbbllx{100bp}%
          \@p@@sbblly{100bp}%
          \@p@@sbburx{200bp}%
          \@p@@sbbury{200bp}%
                \count203=\@p@sbburx
                \count204=\@p@sbbury
                \advance\count203 by -\@p@sbbllx
                \advance\count204 by -\@p@sbblly
                \edef\@bbw{\number\count203}%
                \edef\@bbh{\number\count204}%
          \compute@sizes
          \epsfig@@draft
       \fi
}%
\def\psfig@method{%
        \compute@bb
        \ifepsfbbfound
          \compute@sizes
          \psfig@start
          \vbox to \@p@srheight sp{\hbox to \@p@srwidth 
            sp{\hss}\vss\psfig@end}%
        \else
           \epsfig@draft
        \fi
}%
\def\epsfig@draft{\compute@bb\compute@sizes\epsfig@@draft}%
\def\epsfig@@draft{%
\typeout{<(draft only) \@p@sfilefinal>}%
\if@draftbox
        \hbox{{\fboxsep0pt\fbox{\vbox to \@p@srheight sp{%
        \vss\hbox to \@p@srwidth sp{ \hss 
           \expandafter\Literally\@p@sfilefinal\@nil
                          \hss }\vss
        }}}}%
\else
        \vbox to \@p@srheight sp{%
        \vss\hbox to \@p@srwidth sp{\hss}\vss}%
\fi
}%
\def\Literally#1\@nil{{\tt\graphic@verb{#1}}}
\psfigdriver{dvips}%
\epsfigRestoreAt

\begin{document}
\begin{frontmatter}

\title{Relative chaos in gravitating systems with massive centre}
\author[sussex,technion]{A.A. El-Zant} 
\author[sussex,yerevan]{V.G. Gurzadyan}
\address[sussex]{Astronomy Centre, University of Sussex, Brighton, BN1 9QH, UK}
\address[technion]{Physics Department, Technion --- Israel Institute of Technology,
 Haifa 32000, Israel}
\address[yerevan]{Department of Theoretical Physics, Yerevan
Physics Institute, Yerevan, Armenia}

\begin{abstract}
 
A geometric method, namely the Ricci curvature criterion, is used to study
the relative instability of evolving $N$-body gravitating systems with 
massive centres.
The performed numerical experiments show that the Ricci curvature decreases 
with the increase in central concentration, thus indicating an 
increasing instability for systems with massive centre --- either
in the form of a point mass or a core formed during the evolution of single 
and multimass systems. Simulations of such systems do indeed suggest that they 
have faster gravothermal evolution than less concentrated ones.
\end{abstract}

\end{frontmatter}

\section{Introduction}

It has been known for sometime~\cite{Gurzadyan and Savvidy1,Gurzadyan and Savvidy2} that gravitational systems 
display 
dynamical behaviour that is closer to that of hyperbolic  systems than to near integrable ones. 
In particular, it was
shown that  spherical systems display negative configuration space curvature for the majority 
of initial conditions. 
  As is well known~\cite{Pesin}, 
the behaviour
of hyperbolic dynamical systems is very different from that of near integrable ones.  It follows therefore
that the study of
the stability properties of $N$-body gravitating systems may throw light on 
 basic dynamical  properties of star clusters and galaxies which may not be 
recovered by standard 
techniques of galactic dynamics, which assume that large spherical $N$-body 
gravitational
systems can be treated as integrable (perhaps with the inclusion of {\em additive}
noise to model discreteness effects --- even this however is usually neglected for systems 
with large standard two body relaxation times)~\cite{BT}.

A first step in the study of chaotic
behaviour in numerical models is obviously to find proper descriptors of such behaviour 
and to determine what they tell us about a
given system. The difficulties of the interpretation of computer results concerning $N$-body models are known 
since the pioneering numerical
work of Richard Miller \cite{Miller}. 
The rigorous study of statistical properties of many dimensional non-linear systems, such as
the $N$-body gravitating systems, is also associated with difficulties concerning the 
application of certain criteria and methods.
For example, principal difficulties are associated with the use of Lyapunov exponents, 
given the fact that they are discontinuous
functions of bifurcation parameters \cite{Ruelle}, 
and due to their discontinuity with respect to the computer-created images of
$N$-body systems~\cite{Gurzadyan and Kocharyan4}. 
Lyapunov exponents are also asymptotic quantities which are not particularly
suitable for the study for the local (in time) chaotic behaviour.  of $N$-body gravitational systems.

In the present study of the stability properties of the phase space trajectories of certain $N$-body
configurations we use the geometric approach, well known in the theory of dynamical systems \cite{Arnold}. 
In physical problems
these methods have been introduced originally by Krylov \cite{Krylov}. In stellar dynamics they have been 
first used by Gurzadyan
\& Savvidy \cite{Gurzadyan and Savvidy1,Gurzadyan and Savvidy2} (for further reviews see \cite{GurzadyanP}).
  In the aforementioned work the  
sign indefiniteness of configuration space curvature of $N$-body
gravitating systems was also shown --- meaning that, in general, these systems are not, strictly speaking,
 Anosov systems. 
The latter fact dictates the necessity of
searching for more weak but checkable criteria in characterising the statistical properties 
of $N$-body trajectories.  One
possible way is to find criteria of {\it relative chaos}, i.e. conditions 
which should describe the instability properties of a
given system with respect to another compatible system. 
Such a criterion appears to be one involving the so-called Ricci
curvature, which we will use in the present study. 

The question we are interested in here 
is {\it the effect of the regular field produced by a central mass concentration
on  the stability properties
of an evolving spherical $N$-body system.} 
The massive centre in real star clusters or galaxies can appear as a massive black hole
or central dense stellar core (resulting from core 
collapse or  mass segregation resulting from energy equipartition in multimass systems). 
However, various physical processes directly associated with
 both these phenomena (loss cone effects, for example) and 
which are crucial in the vicinity of the centre, 
are out of the scope of our analysis.

Motivation for our study stems in part from a series of results where  a 
central mass 
concentration has been found to affect the dynamics of gravitational 
systems in an important way. 
The destabilizing role of a regular  central field has been observed 
by Gurzadyan \& Kocharyan 
\cite{Gurzadyan and Kocharyan8} using the Ricci curvature as a chaos indicator;
in that study only small static systems were investigated, while in the present paper
we will investigate systems with far larger number of particles
which start near detailed  {\em dynamical}
equilibria and which may evolve in time.   
 Even before it has been noticed in $N$-body simulations that 
asymmetric 
systems quickly evolve towards more isotropic shapes when a significant  central mass 
is present (\cite{Norman et al.1}; see also \cite{Norman et al.2})
an effect due to the destabilization of the main periodic orbits. 
The increase in the rate of relaxation for systems with
massive centre (black hole) has been studied also by 
Rauch \& Tremaine \cite{Rauch and Tremaine} 
where the effect of resonant relaxation was investigated.
It was also shown that the 
central mass makes  spherical system more unstable with respect to the
gravothermal catastrophe \cite{Gurzadyan et al.}; that result has been
obtained by means of methods of catastrophe theory.   
The stability of systems with a central point mass has been studied
also by \cite{Smith et al.}; it was shown that not only the neighbouring
particles have role in the stability, but also the mean field.

The content of this paper is as follows.
In Section 2 we describe the geometric criterion for 
intstability used in this 
paper, in Section 3 we discuss the numerical method, 
Section 4 presents the results which
include simulations of single mass systems without initial central 
concentrations (Section 4.1) and
systems with initial central concentrations or with different particle masses 
(Section 4.2). The results are discussed in Section 5.

\section{Ricci curvature criterion}

For an introduction to the basic ideas behind the use of
 geometric methods in the theory of dynamical systems the reader is 
referred to reference \cite{Arnold}. 
Here we briefly present the  
basic concepts needed in the interpretation of the numerical results. 
The Ricci curvature as a criterion for relative chaos has been 
introduced by Gurzadyan and Kocharyan in \cite{Gurzadyan and Kocharyan7}, 
where a kind of
classification of various stellar configurations by their degree of chaos
had been also achieved. 
The criterion is based on the computation of the Ricci curvature in the
direction of the  geodesic velocity ${\bf u}$ of the phase space of the system,
and is defined as follows \cite{Eisenhart,Grommol et al.}:
$$
r_{\bf u}(s)= \frac{{\bf Ric}({\bf u,u})}{ \parallel {\bf u} \parallel^2}
=\sum_{\mu=1}^{3N-1}
k_{{\bf n}_{\mu},{\bf u} }(s),
$$
where ${\bf Ric}$ is the Ricci tensor, $s$ is the geodesic time,
$k_{\bf n,u}$ is the two-dimensional curvature,
and the deviation vectors are chosen as
 ${\bf n}_{\mu}\bot {\bf u}$ and ${\bf n}_{\mu}\bot {\bf n}_{\nu}$ for all $\nu \neq \mu$.

This form of the Ricci curvature  arises while
averaging the Jacobi equation ---  which links the dynamics
of the flow with the geometry of the phase space. One then obtains 
$$
\frac{d^2 \parallel {\bf n} \parallel}{ds^2} =  
- \frac{1}{3N} r_{\bf u} +  \langle \parallel \nabla_{\bf u} {\bf n}
\parallel^2 \rangle.
$$ 
Then, 
the {\it criterion of relative instability} reads:
among two systems with  $r_1$ and $r_2$ within an
interval of affine parameter $s$ - $[0, s^*]$, the system with smaller 
negative $r$ should typically be more unstable, i.e. unstable with
higher probability, where 
$$
r= \frac{1}{3N} \inf_{0\leq s\leq s_*}\left[r_{\bf u} (s) \right],
\qquad r_1 < 0; r_1 < r_2;
$$
Certain loss of information on 
configurations of the system is obvious, however the Ricci
curvature does contain  
information on  typical systems, as distinct to the
scalar curvature which does not contain it. 

 The explicit form
of the Ricci tensor readily follows from the Riemann tensor
 is given by \cite{Gurzadyan and Savvidy1}  
$$
 Ric_{\lambda\rho} = -(1/2W)
\left[\Delta W g_{\lambda\rho}+(3N-2)W_{\lambda\rho} \right]
$$
$$
+ (3/4W^2) \left[(3N-2)W_{\lambda}W_{\rho}\right]-
\left[\frac{3}{4W^2}-\frac{(3N-1)}{4W^2}\right]g_{\lambda\rho}
\parallel  dW \parallel^2.
$$
where 
$$
W=E-V(q)=E-\sum_{b<a}^{N}Gm_am_b/r_{ab},
$$
and $m_i$ are the masses of the particles of the $N$-body system.

\section{Numerical method and choice of  system}

The numerical $N$-body routine used in this study is the NBODY2 code developed by Sverre Aarseth~\cite{Aarseth} 
and kindly made available by him.  
This is a direct summation code which uses individual time-steps for each
particle in the simulation  and speeds up the force calculation by using a
 neighbour scheme which separates
 the force calculations for neighbouring particles from  those further off.  These improvements take into
 account the very different natural times ($\sim 1/\sqrt{\rho}$, $\rho$ being the local density) in a gravitational system and the
fact that, at a given point, the {\em irregular} force due to nearby neighbours varies much faster than the {\em regular} force
due to particles further off.  For highly inhomogeneous centrally concentrated systems these refinements can lead to a 
considerable increase in efficiency while essentially  maintaining the high accuracy of direct summation method. 
The errors in the calculations (as measured by energy conservation) are controlled by an accuracy parameter $\eta$ which
determines the size of the integration time-steps.  These errors are constant for values of $\eta$ below 0.01 and increase
 as $\eta^{2}$. We have found that a value of $\eta_{irr}=0.02$ (controlling the irregular
time-step)  and $\eta_{reg}=0.04$ (for the regular time-step) gave reasonable results while maintaining efficient
running of the code. 
For $N \leq 1000$ the maximum number of particles allowed in the neighbour 
sphere was $10 + \sqrt{N}$ while for larger $N$ it was taken to be
$\left(\sqrt{\frac{\eta_{reg}}{\eta_{irr}}} \frac{N}{4} \right)^{3/4}$. 

The systems are taken to start from Plummer 
sphere configurations with the initial conditions 
determined by the method described in \cite{Aarseth et al.}.
Using the units of Heggie \& Mathieu \cite{Heggie and Mathieu} 
(by setting the total mass and the gravitational constant 
to unity while keeping the total energy of
the system at $E= -0.25$) the Plummer model is defined completely:
the virial  radius is fixed to
unity and the mean virial equilibrium crossing time is equal to $2 \sqrt{2}$ time units.

The NBODY2 code is not designed to work with zero softening \cite{Aarseth},
therefore a small softening, again of  Plummer form, has been added 
to avoid numerical overflow  due to
close encounters and large angle scatterings. The 
 softening parameter was
taken as $\epsilon=2/N$  which scales as the ratio of the 
minimum to maximum impact parameters of standard relaxation theory \cite{Giersz and Heggie}.

 Since  the Ricci curvature determines the {\em average} 
or typical instability response
 to random perturbation, it requires  not too large dispersions.
 It is therefore
necessary to remove the larger contributions due to close encounters.
In \cite{El-Zant1} the Ricci curvature 
time series  corresponding to a given simulation have been averaged over small time intervals 
while filtering out terms that were much larger than the typical value.
Here instead we simply not include 
contributions to the Ricci curvature 
due to interactions of particles within a (softened)
radius of 0.05 units from each other
($5 \% $ of the virial radius). The typical fraction of operations excluded by
this procedure is about a few in ten thousand.

\section{Results}

\subsection{Variation of the Ricci curvature as a result of core contraction}

\begin{figure}
\hspace{1cm}
\epsfig{file=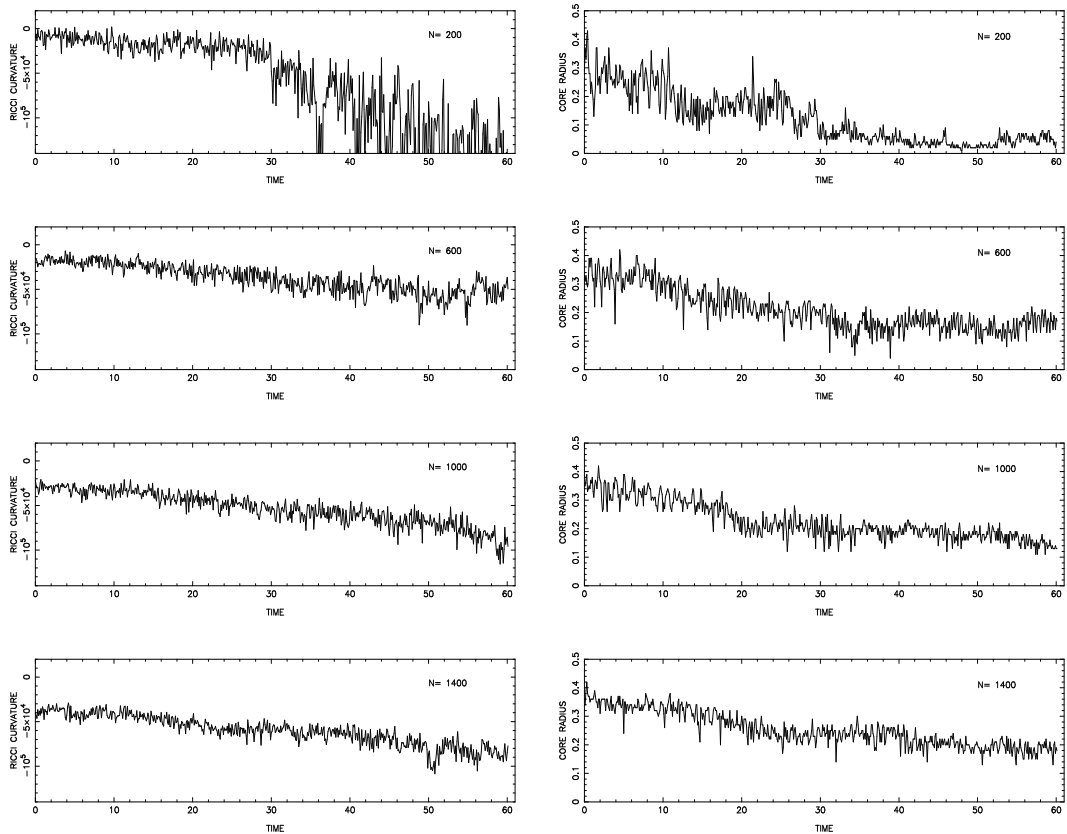,width=14.0cm,height=11.0cm,angle=-90}
\caption{\label{coent}
Evolution of the core radius and Ricci curvature for different realizations of
the same Plummer model}
\end{figure}

The diagrams on the left hand side of Fig.~\ref{coent}  
show the time evolution of the Ricci curvature, as a function of the crossing time,
for systems starting from different $N$-body realizations of the same Plummer  model.
  The uppermost plot corresponds to the evolution of $200$ particle system,
while the following plots correspond to the evolution of systems consisting
of $600$, $1000$ and $1400$ particles respectively. Clearly 
the Ricci curvature is {\em always} negative and becomes more so
as the systems evolve. This can be understood readily. 

Since the pioneering work of Antonov \cite{Antonov} and Lynden-Bell and 
Wood \cite{Lynden-Bell and Wood}, it is well known that 
gravitating systems in virial equilibrium 
have no entropy  maxima. Instead they can continually evolve 
towards higher entropy states
characterised by the appearance of
progressively tighter cores surrounded by  a diffuse halo. 

It  is  therefore natural to take the evolution of the core radius as
an indicator of the evolutionary state of a spherical system.
Plummer spheres do indeed tend towards more concentrated configurations as they 
evolve, we must therefore assume that these higher entropy  states are characterised by smaller values of
the Ricci curvature, i.e. with higher dynamical entropy states. 
Note that here a correlation does exist between the dynamic entropy given
by Ricci curvature and the thermodynamic one, which is not the case for
any physical states of gravitating configurations because of their non-compact
phase space.

NBODY2 contains a routine for finding the core radius of a spherical system using the prescription
of Casertano and Hut \cite{Casertano + Hut}.
Particles inside a given radius from the origin (usually the half mass radius) are selected. For each of 
these  particles a list is made of the six nearest particles. The ``mass density'' for
particle $i$ is then defined as $\rho_{i} = 3 M_{5}/(4 \pi r^{3}_{6})$ 
where $r_{6}$ is the
distance to the particle farthest away from particle $i$ in the list and $M_{5}$ is the sum of the masses
of the five other particles. The density centre of the system 
\begin{equation}
{\bf r}_{d}= \sum {\bf r}_{i} \rho_{i}/
\sum \rho_{i}
\end{equation} 
is then calculated. With the help of this quantity the core radius is then defined to be
\begin{equation}
R_{c}=\sqrt { \frac{ \sum \parallel {\bf r}_{i} - 
{\bf r}_{d} \parallel^{2} \rho^{2}_{i} } { \sum \rho^{2}_{i} } }.
\label{sphex:core}
\end{equation}

The plots on the right hand side  of Fig.~\ref{coent}  show the evolution of the core 
radius as calculated from Eq.~(\ref{sphex:core}) 
for the runs which correspond to the Ricci 
curvature plots adjacent to them. 
These diagrams clearly show  that {\it the evolution of the Ricci
curvature properly correlates with the evolution of the core radius},
in the sense that they 
both decrease together, thus confirming the conjecture outlined above.
Only the scaling in the Ricci curvature undergoes changes with 
the number of particles, but not its behaviour.

\subsection{Massive centre and multimass systems}
\label{sphex:plumcdm}

As we
have seen above, as the Plummer sphere models become  more 
concentrated their dynamical entropy, as measured by the negative Ricci curvature, increases.
Hence if there is any correlation  between this quantity and the rate
of evolution of a system towards a tighter core, then we expect centrally 
concentrated systems to evolve faster, thus outlining the  self amplifying
nature of gravothermal instability.  
\begin{figure}
\hspace{1cm}
\begin{center}
\epsfig{file=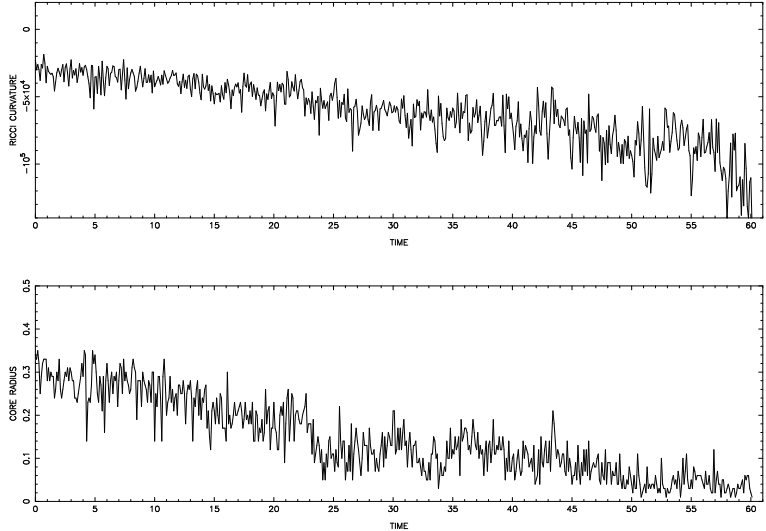,width=10.0cm,height=7.0cm,angle=-90}
\end{center}
\caption{\label{cenent}
 Evolution of core radius and Ricci curvature for a $N=1000$ system in which of one of
the particles, initially located at the centre, is ten times more massive than 
each of the remaining 999 particles}
\end{figure}

To see if the above argument is valid  we have conducted simulations in
which one of the particles, which is initially
 placed at the centre of the system, 
has a mass that is much larger than the other remaining $999$ particles.
The mass of the central particle was taken to be either ten, fifty or a 
hundred times the mass of the other particles in the simulation. 
Other  system parameters like the total mass, energy, size etc were 
kept constant. The Plummer model is rescaled so as to start from virial
equilibrium including the heavier particle. The systems stay near virial 
equilibrium throughout the simulations --- in the case when the central particle
is ten times more massive than the other particles, the virial ratio 
$T/W$ does not vary by more than $2\%$. This  suggests that our 
models start from near a detailed dynamical equilibrium and that initial 
departure from such an equilibrium would not play a drastic role in 
determining the subsequent evolution.
Because of its significantly larger mass the particle remains essentially
within the inner $0.3 \%$ of  the  (tidal) radius of the system even
for the case when it is only ten times the mass of the other particles.

We show the corresponding plots for the Ricci curvature and core radius 
time evolutions in Fig.~\ref{cenent}, which should be compared with the $N= 1000$ plots
of Fig.~\ref{coent}. It is evident that the evolution of the core
radius is much faster for the system with a central massive particle confirming
that such system is more unstable. Thus,  again {\it the 
evolution of the Ricci curvature correlates with the evolution    
towards a denser core}. This result is perhaps not that surprising since 
core contraction is related to the negative specific heat of open gravitational
systems --- which means that a subsystem losing energy {\em increases} its kinetic
energy. This process is amplified when the velocity gradient in the mean field
is steep,
which is precisely what happens when there is a central density cusp or concentration.

\begin{table}
\begin{center}
\begin{tabular}{c|c|c}
\multicolumn{3}{c}{ }\\
\hline
$M_{p}/\bar{M}$          & $\tau$ & $\epsilon_{10}$\\ 
\hline
10                       & 0.29        & 0.01 \\
\hline
50                       & 0.22        & 0.06\\
\hline
\hspace{0.12in} 100 \hspace{0.12in} & 
\hspace{0.12in}  0.13 \hspace{0.12in} & \hspace{0.12in} 0.18 
\hspace{0.12in}
\end{tabular}
\end{center}
\caption{\label{sphex:tabC} Time-scales (in crossing times)
  derived from 
the values of scalar curvature and averaged over  
ten outputs during the first crossing time for $N=1000$ 
systems in which one of the particles, originally located at the centre, has
a mass $M_{p}$ that 
either 10, 50 or a 100 times more massive than the mass of each of the remaining  999 particles of mass $\bar{M}$.
The third column estimates the RMS error in the
ten outputs}. 
\end{table}
\vspace{0.1in}

Similar behaviour was found for the system with higher masses for the central particles. 
However, for $N=1000$, the data for these systems were found to be more noisy. Both the Ricci curvature
and the core radius calculations were found to be dominated by the position of the central
massive body and by close encounters of this particle with the surrounding ones.
Also, 
 due to the continuing tightening of the core and the formation of hard binaries,
numerical difficulties were encountered since the NBODY2 contains no provisions for
handling such effects. The simulations had to be stopped because they had slowed down
dramatically after about twenty two crossing times. It is interesting to note that, in
the case when the central particle has a mass equal to a hundred times that
of the other particles, the core radius had already contracted to $0.04$ units after
a few ($\sim 3$) crossing times. Table~\ref{sphex:tabC} shows the 
characteristic time-scales (in units of the crossing time) averaged
over the first ten outputs (corresponding to an interval of about a 
crossing  
time) for the different models. These are derived from the values of the 
scalar curvature (which are less noisy). The use of scalar curvature is motivated by the analytical
proof in references~\cite{Gurzadyan and Savvidy1,Gurzadyan and Savvidy2} that the
relaxation time-scale of spherical $N$-body systems is determined by the scalar
curvature.

\begin{figure} 
\hspace{1cm} 
\begin{center} 
\epsfig{file=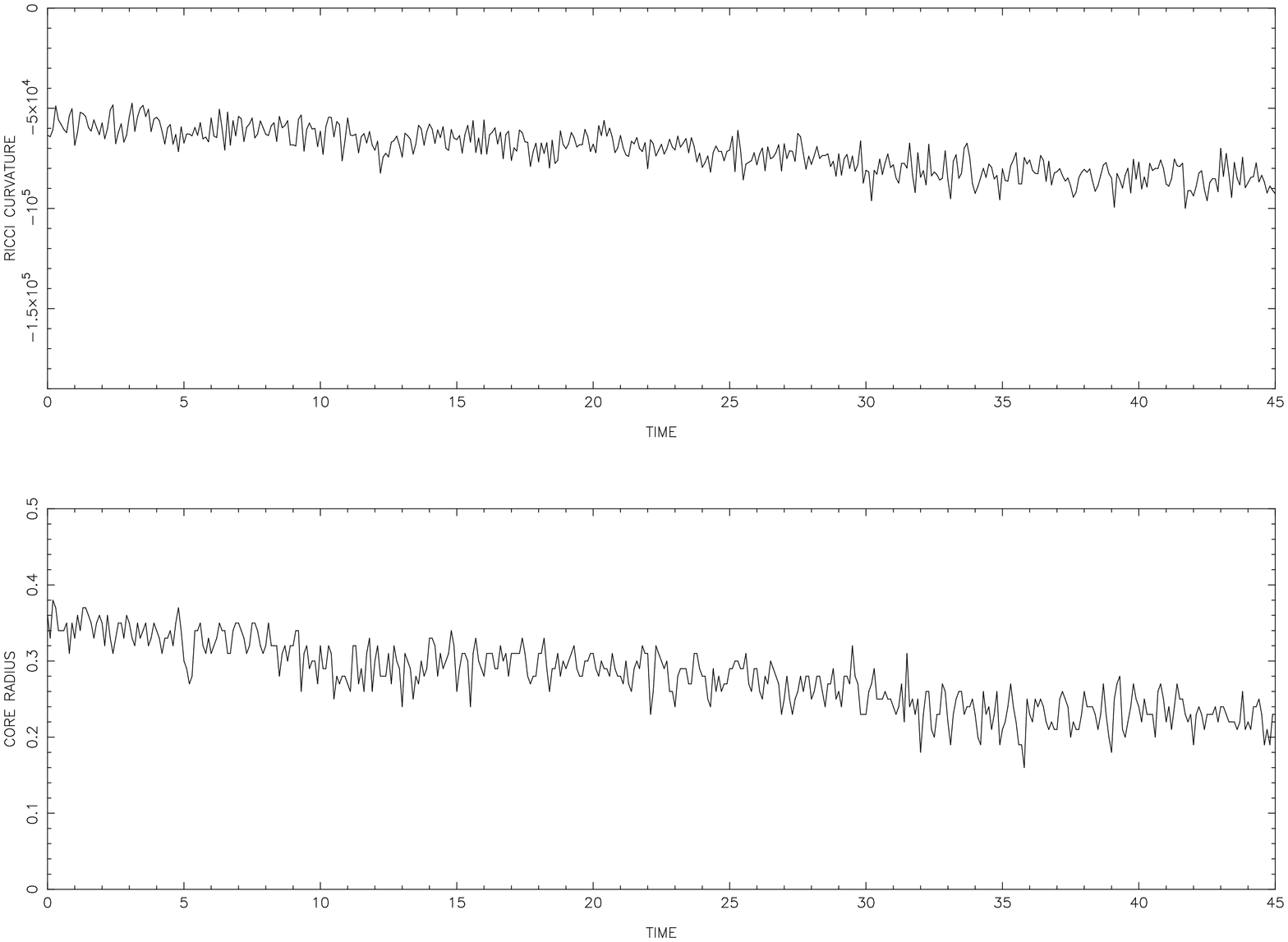,width=11.0cm,height=7.0cm,angle=-90} 

\vspace{0.3cm}

\epsfig{file=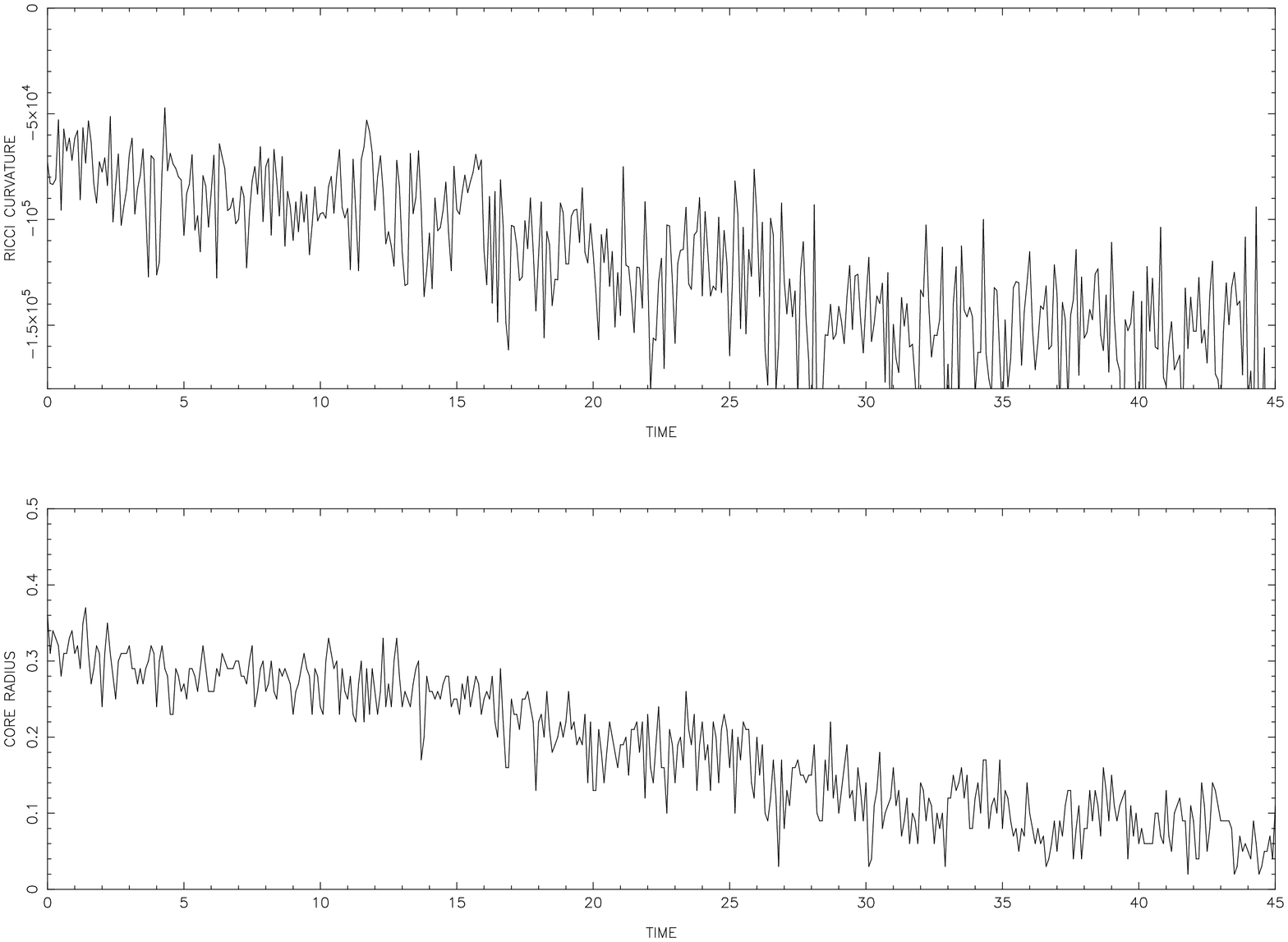,width=11.0cm,height=7.0cm,angle=-90} 

\vspace{0.3cm}

\epsfig{file=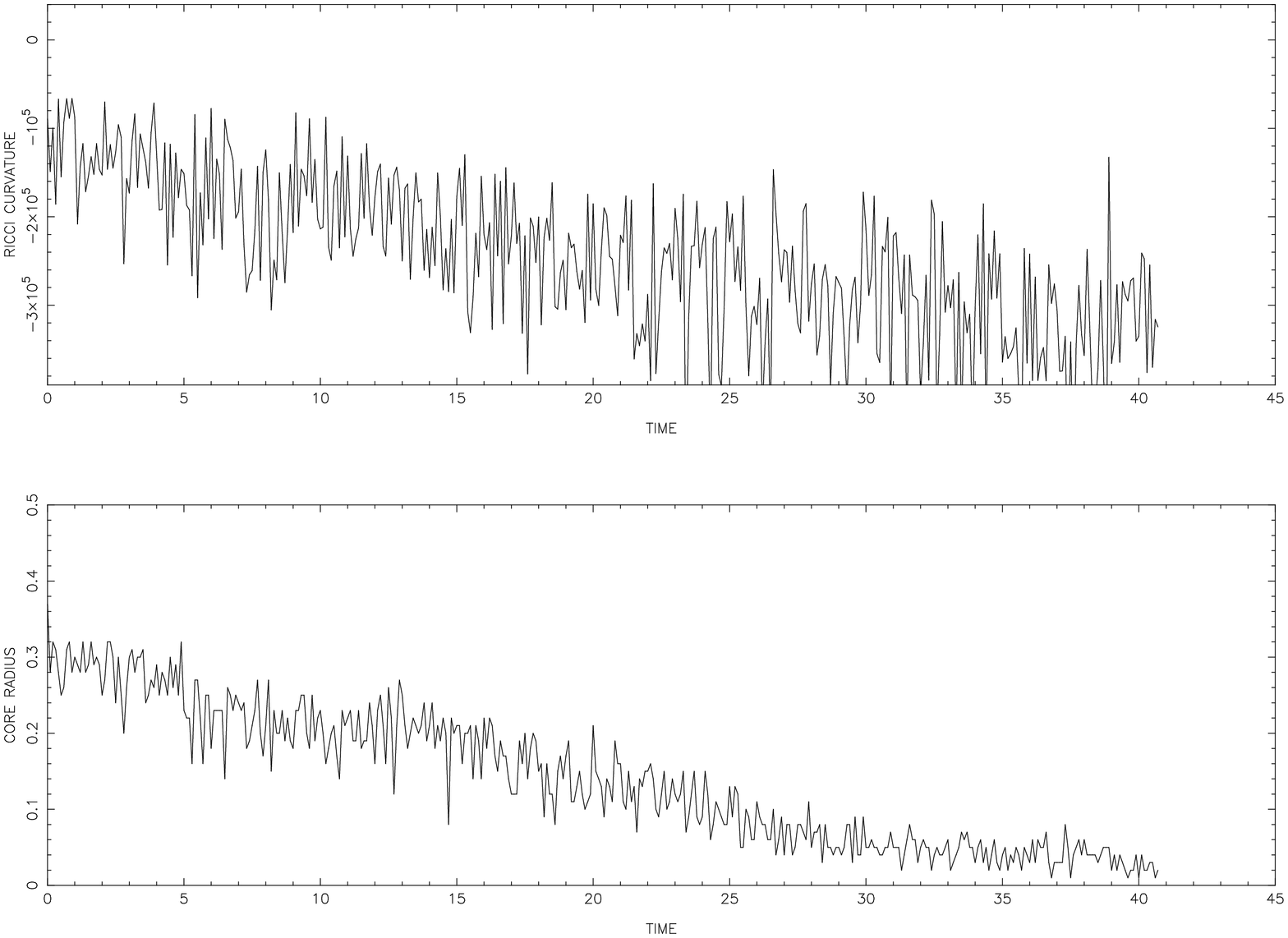,width=11.0cm,height=7.0cm,angle=-90}
\end{center} 
\caption{\label{run1sr}
 Evolution of core radius and Ricci curvature for  $N=2000$ systems in which of one of
the particles, initially located at the centre, is (from top to bottom) is 10,
50 and  100 times more massive than each of the remaining 1999 particles}
\end{figure}

\begin{figure} 
\hspace{1cm} 
\begin{center} 
\epsfig{file=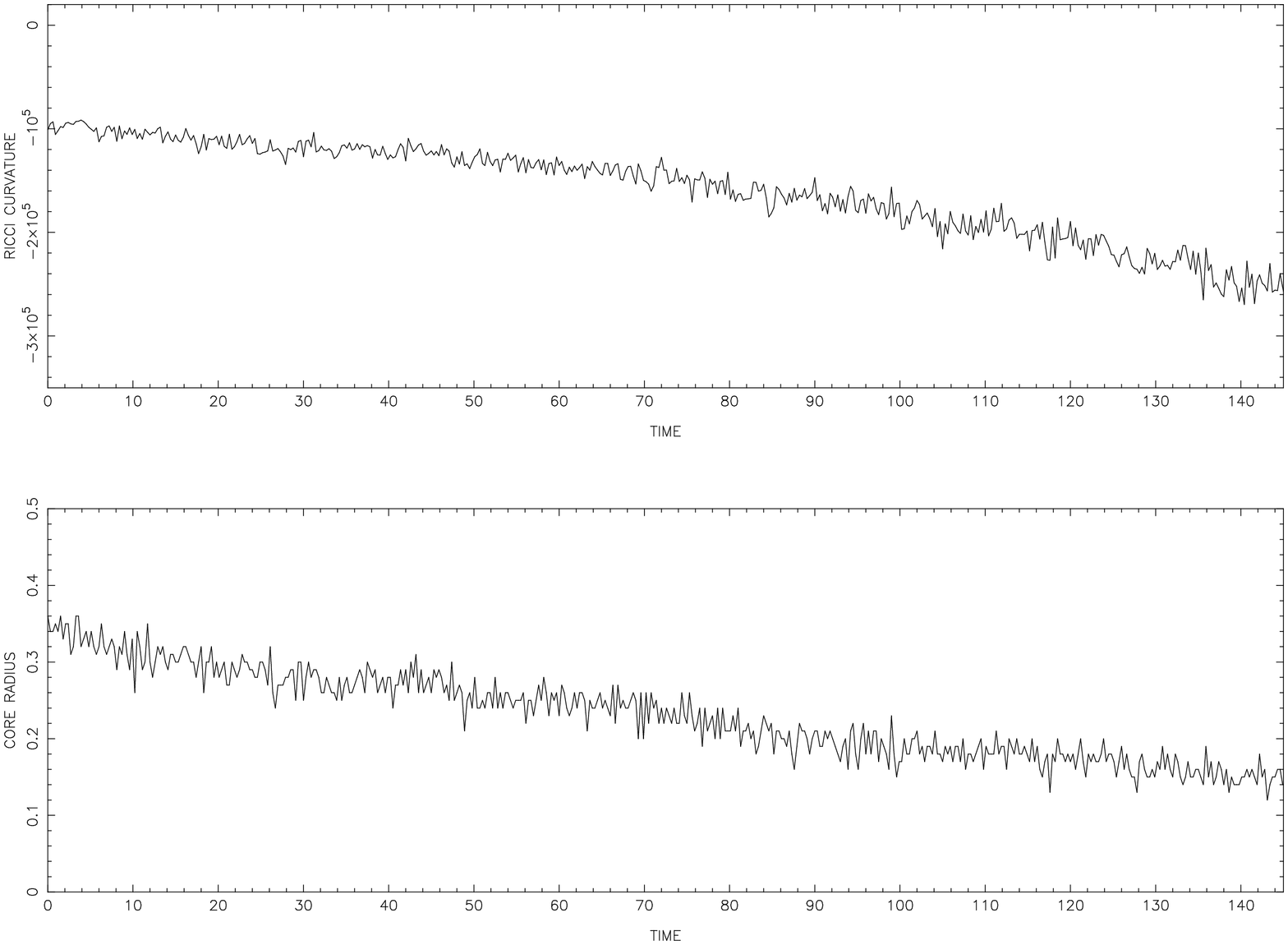,width=11.0cm,height=7.0cm,angle=-90} 

\vspace{0.3cm}

\epsfig{file=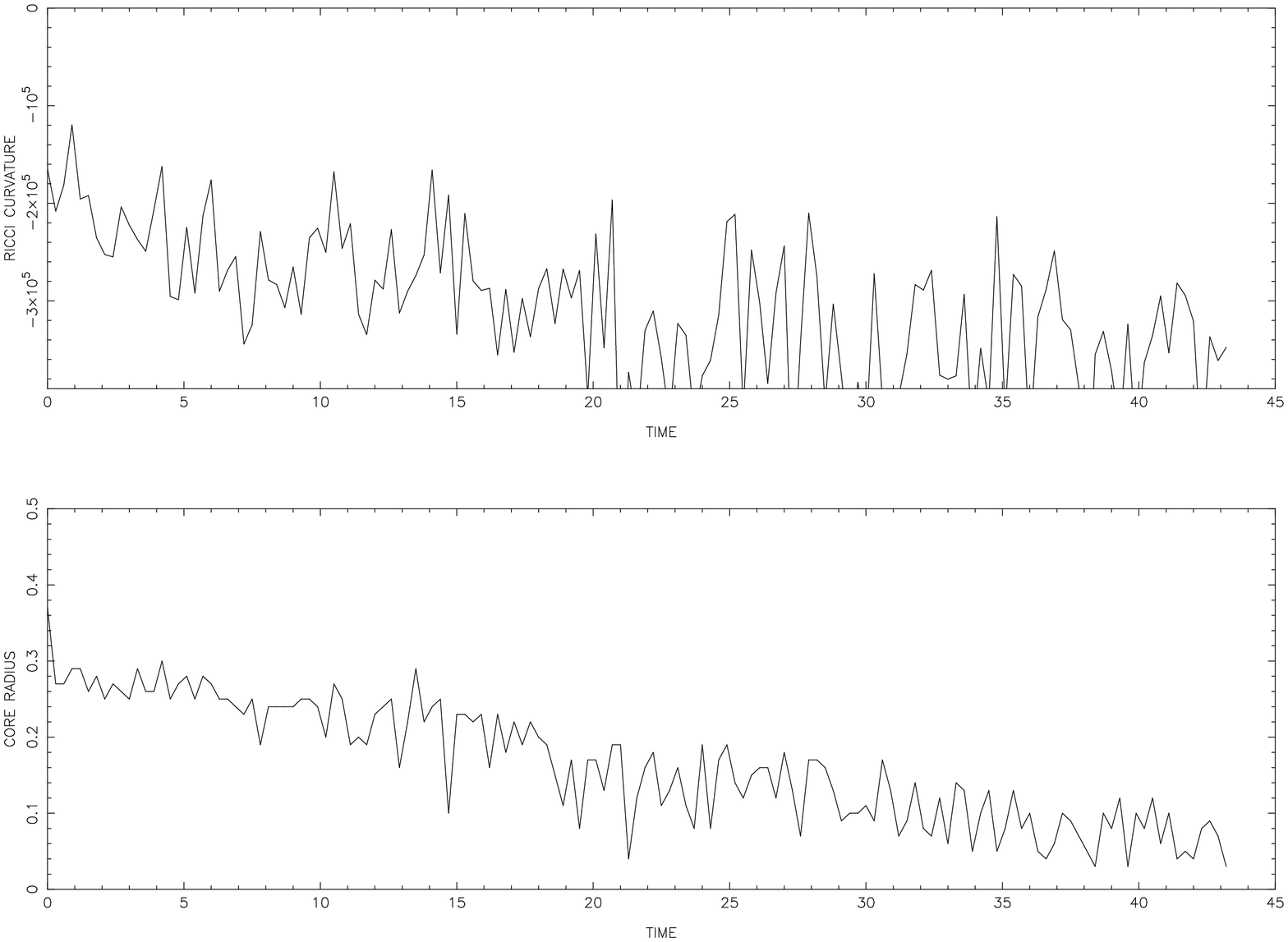,width=11.0cm,height=7.0cm,angle=-90} 
\end{center}

\caption{\label{run1r}
 Evolution of core radius and Ricci curvature for  $N=3500$ systems when  one of
the particles, initially located at the centre, is   10 (top) or 
150  hundred times  more massive than each of the remaining 3499 particles}
\end{figure}

Fig.~\ref{run1sr}  shows the behaviour of the Ricci curvature and core radii for systems consisting of two thousand 
particles when the central particle is ten, fifty and one hundred times more massive than the remaining particles. 
The estimate of the core radius was again found to be very noisy if the central particle is included.
Moreover this estimate then reflected the large  mass concentrated in the  central 
particle  rather  than the overall contraction of the core due to the dynamical evolution.
 For these reasons, 
the contribution of this particle was removed when calculating the aforementioned 
quantity. This type  of filtering was not applied  in 
the calculation of the Ricci curvature (which is a characteristic of the self consistent dynamics). 
As a result the time 
evolution of this quantity contains larger fluctuations.
Again, the larger the central mass the faster the evolution. 
While the presence of a central particle only ten times more 
massive than the remaining ones (a perturbation of $0.5 \%$) was not found to have a marked effect 
on the evolution, systems with central masses of $1/40$ and $1/20$
 the total mass of the system evolved significantly more rapidly.
These results were also confirmed for systems with $N=3500$. The results for these systems are shown 
in Fig.~\ref{run1r} Once more we see that when the mass of the central particle is sufficiently small,
the evolution is relatively slow. It appears that (at least for the $N=1000,2000$ and $3500$) 
a central mass of $\sim 1\%$ is sufficient to significantly accellerate the evolution of spherical
gravitational systems.

\begin{figure}
\hspace{1cm}
\begin{center}
\epsfig{file=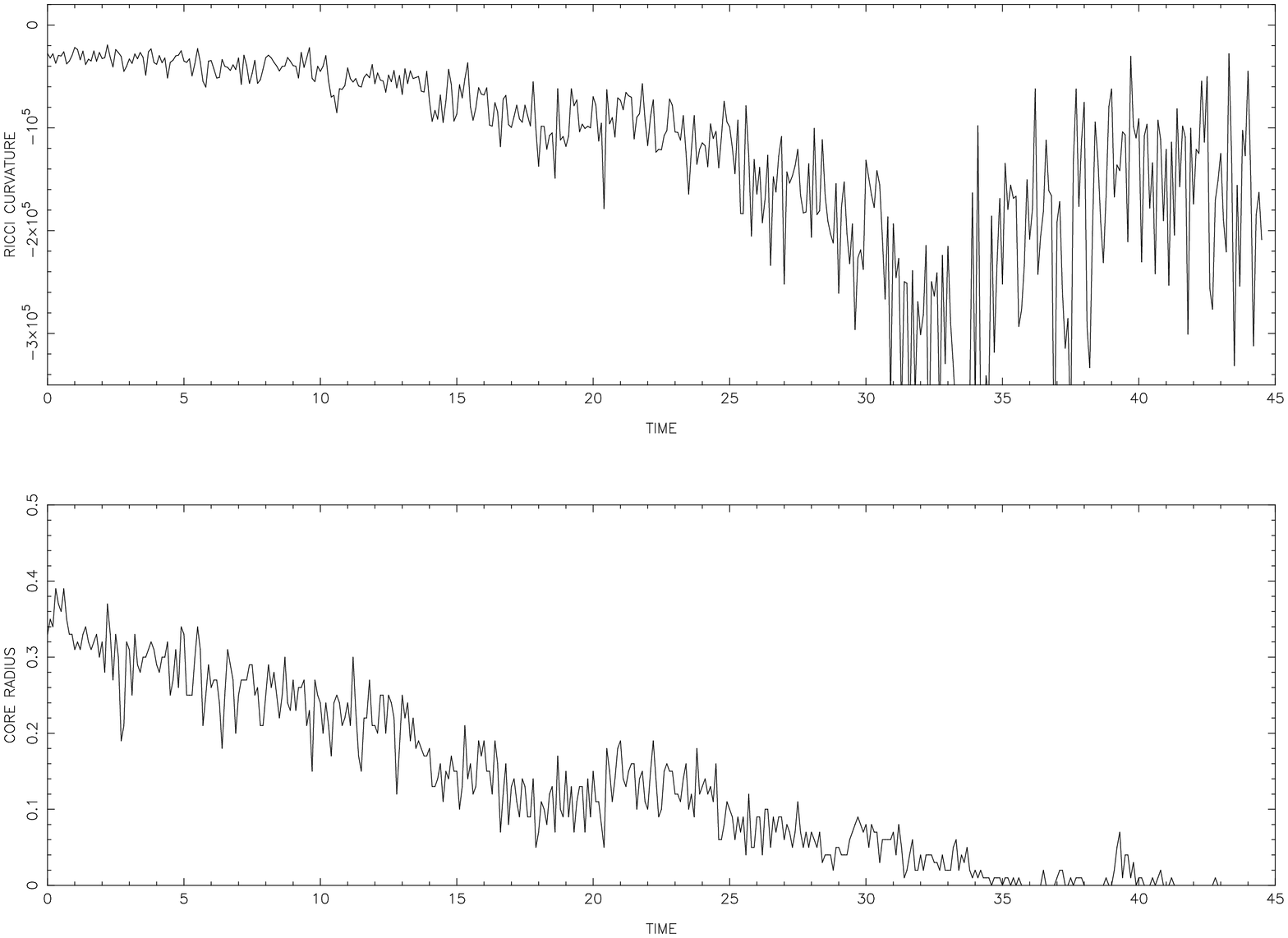,width=10.0cm,height=7.0cm,angle=-90}
\end{center}
\caption{\label{curdm}
 Evolution of core radius and Ricci curvature for $N=1000$ multimass system with Salpeter
mass function and in which the  highest mass particle is ten times more massive than the 
lowest mass one}
\end{figure}

\begin{figure}
\hspace{1cm}
\begin{center}
\epsfig{file=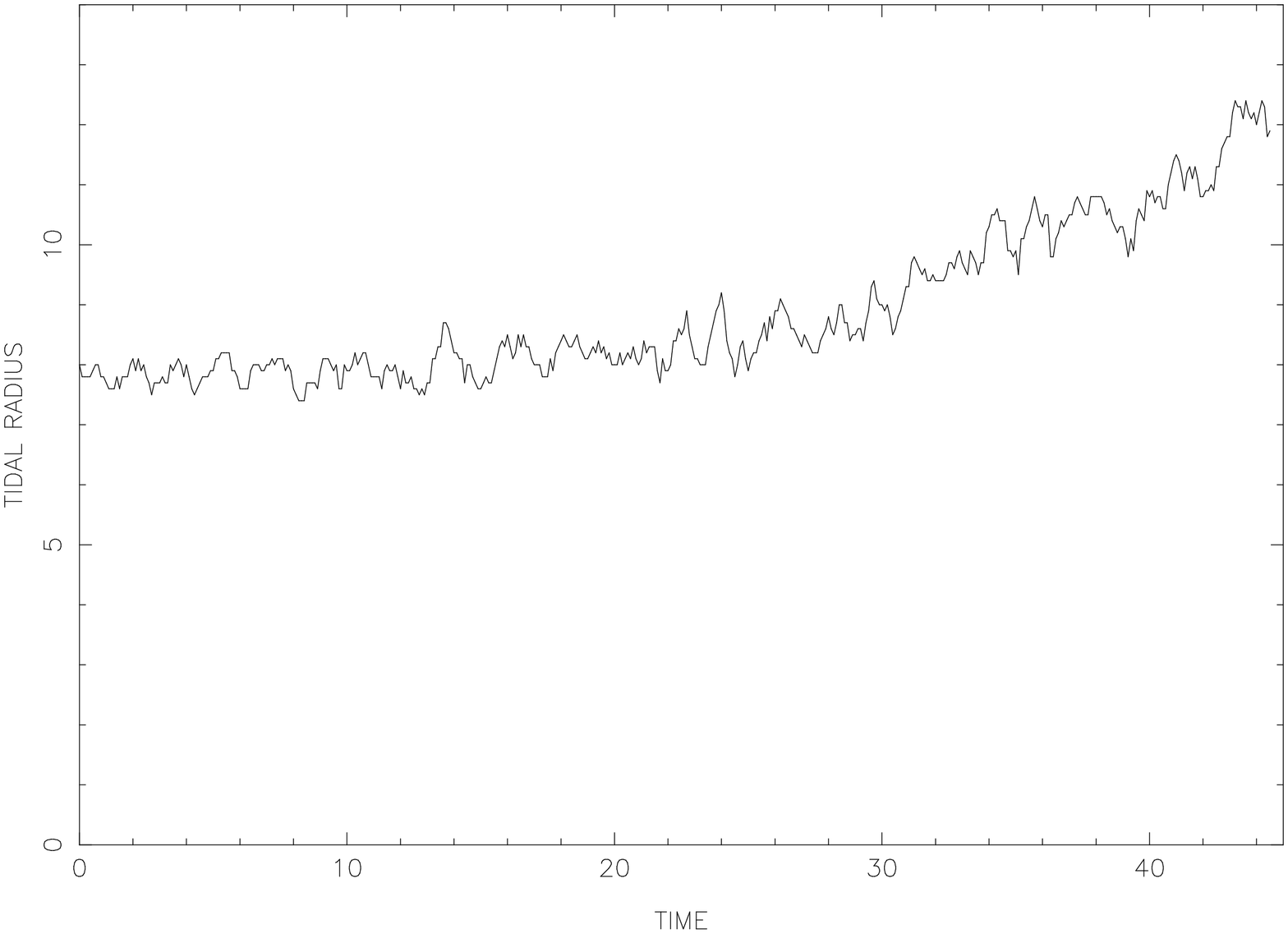,width=10.0cm,height=4.0cm,angle=-90}
\end{center}
\caption{\label{tidadif}
 Evolution of tidal radius of the system considered in Fig.~\protect\ref{curdm}}
\end{figure}

In a system consisting of particles with different masses, interactions between 
particles leading to equipartition of energy means that more massive particles
tend to lose kinetic energy and reside near the centre. This  leads to
a growth in the central mass concentration and the processes described above will
then take place.  As can be seen from Fig.~\ref{curdm}  this is clearly the case. Actually
the evolution in the multimass system is even much more rapid. This is due to
the fact that here there are two effects  --- the exchange of energy leading to high
central mass and the effect of the central mass itself ---
which reinforce each other and drive the evolution. After about thirty  crossing times
many particles have effectively escaped the system and many more are ejected from the
central areas and, except for the tightening central core, the system becomes more diffuse.
 This is clear from the evolution of the tidal radius shown in Fig.~\ref{tidadif}. 
The core radius becomes very small and only a few particles are enclosed by
it (rendering the calculation of that quantity to be itself inaccurate). 
Hard binaries  start forming and become tighter as the system evolves. These effects
cause the integration to practically stop
 at about forty five crossing times. Before that, the region surrounding the core reexpands as the negative
specific heat is reversed by the energy stored in the binaries.
This  state of the system is actually more
regular than the intermediate core collapse  since many of the escaping particles 
are moving in regular orbits and the outer areas are very diffuse. 
This is one of the reasons for the increase of the Ricci
curvature near the end of the simulation --- the others being that at these later stages
a large contribution to the Ricci curvature is due to the tight central binaries performing  regular two body motion.
At this stage we have reached the limits of the applicability of our geometric
chaos indicator --- an effect due to the non-compactness of the phase 
space.

\begin{figure} 
\hspace{1cm} 
\begin{center} 
\epsfig{file=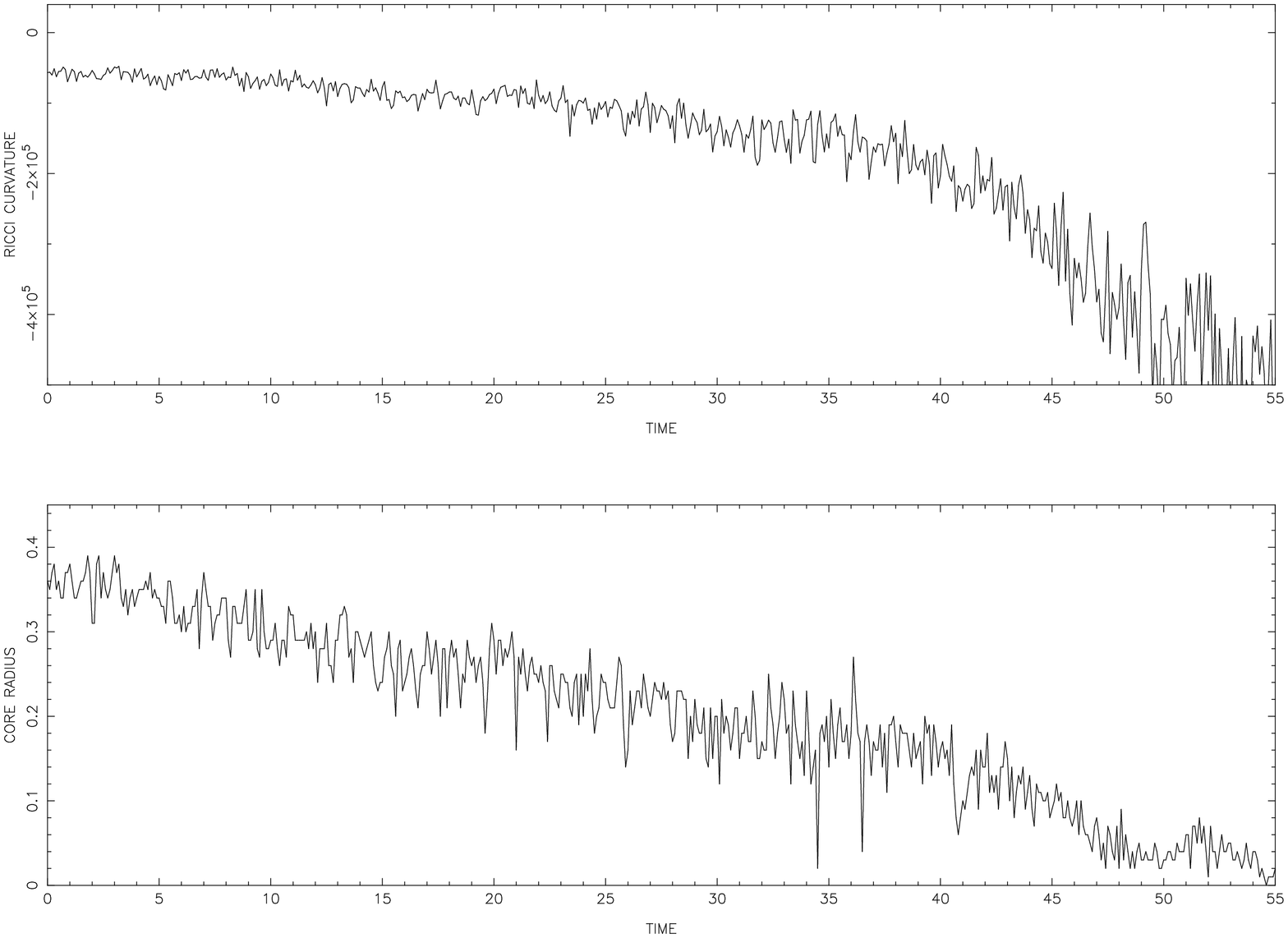,width=11.0cm,height=7.0cm,angle=-90} 

\vspace{0.3cm}

\epsfig{file=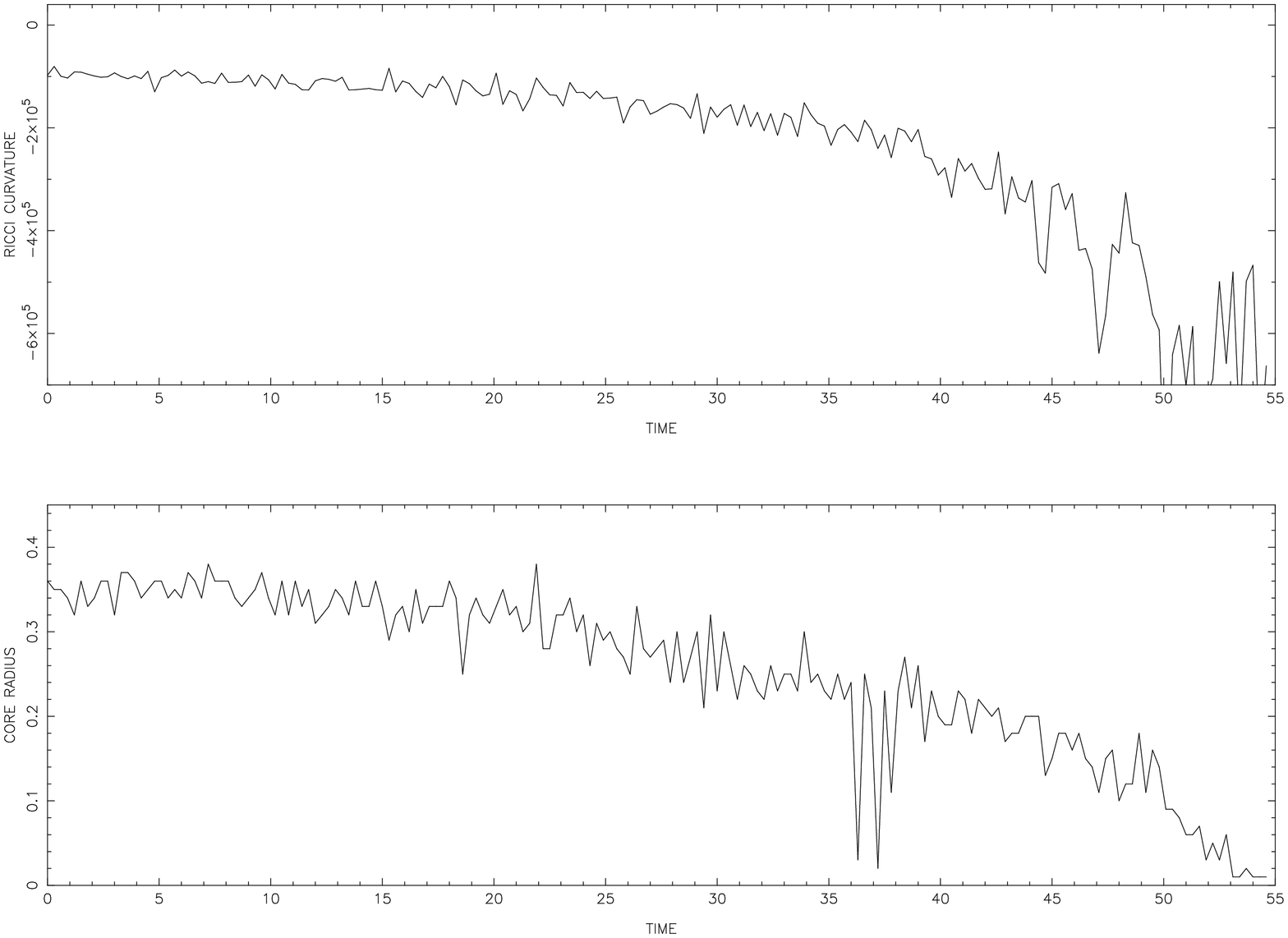,width=11.0cm,height=7.0cm,angle=-90} 

\end{center}

\caption{\label{runmrs}
 Evolution of core radius and Ricci curvature for  $N=2000$ (top) and 
$N=3500$ multimass systems with Salpeter mass functions. The 
most massive particles are 10 and 20 times more massive than the least massive
ones in the $N=2000$ and the $N=3500$ respectively}
\end{figure}

Multimass systems with $N=2000$ and $N=3500$ were also found to evolve faster than their
single mass counterparts with similar global properties.
This can be seen from  Fig.~\ref{runmrs} which shows the the evolution of Ricci curvatures and core radii 
for systems  with Salpeter mass function and where the most massive
particle has a mass ten (for the case when $N=2000$) and twenty (for $N=3500$)
times  larger than the least massive particle (respectively). Again there is a clear general
correlation between the evolution of the two quantities. However, as can be seen, there is some difference 
in the detailed behaviour. This might be expected since the Ricci curvature and the core radius 
are defined in rather different ways. Therefore fluctuations due to the finiteness of the system 
are expected to be different. In particular, the core radius is not a characteristic quantity of 
the dynamics of the system but is a (rather arbitrarily defined) indicator of its central concentration
and depends only on the positions of the particles.

\section{Discussion and conclusions}

In the present study the statistical
 properties of certain $N$-body systems have been analysed, 
particularly the role of the field of a massive centre is revealed. The massive centre
in stellar systems can be imagined either as a central point mass
(massive black hole) or dense stellar core formed during the
gravothermal evolution of the system. Therefore we have considered both
cases, even though one can expect a priori that their resulting 
dynamical effect on the system should be similar (and indeed this turns out to be the case).

The main result of the present study is that
{\it a massive centre leads to an increase in the instability (i.e. mixing) 
properties of evolving N-body gravitating system}. 
This increase in the mixing properties leads to faster gravothermal evolution
for centrally concentrated systems underlining the self amplifying nature of
the gravothermal catastrophe, in which central concentration leads to an 
increase in the temperature gradient which in turn leads to an 
acceleration of the process. Indeed, it was found that  systems with 
central mass concentration, either in the form of a point mass initially
placed at the centre or multimass systems which evolved a central 
concentration because of the accumulation of massive particles near the 
centre, evolved  faster.

One prediction of this study, therefore, is that stellar clusters with initial
central mass concentrations (such as primordial black holes for example) will
evolve faster towards core collapse.  
The faster mixing rate also means that a perturbed system returns to its 
unperturbed state at a faster rate.
This  indicates that, for example, a star cluster with massive centre 
being shocked by Galactic tidal field, will recover
its spherical  shape  faster than a cluster without a central mass.
This problem  is associated with that of the ellipticity of Galactic globular clusters,
 attracting now much attention. The relative rate of evolution  should be proportional
to the values of the Ricci curvature, which by definition describe not
the absolute but relative degree of instability of different systems.  

Since centrally concentrated states of gravitating systems have higher
thermodynamic entropy, it follows that the dynamical entropy 
(which increases with decreasing curvature) is correlated with the 
thermodynamic entropy in gravitational systems (at least for the cases studied).
In previous
studies~\cite{El-Zant1,El-Zant2} it was  shown that  the 
existence of large scale ordered motion also decreased the dynamical entropy --- thus also 
suggesting the above conclusion. In particular, El-Zant \cite{El-Zant1} has 
found that
the dynamical entropy increased during the evolution of  initially ordered 
systems with
macroscopic  (e.g., plasma type collective) instabilities.
It is important to note however that this correllation between thermodynamical and dynamamical
entropy does not hold in the later stages of evolution of $N$-body systems, when effects due to the 
non-compactedness of the phase space are prominent. This is because,
 striclty speaking, the methods we have employed  here are applicable for systems with a 
compact phase space. The dynamics in these later stages are actually more regular as the system 
disintegrates. The peculiar thermodynamic properties of gravitational interaction allows such states 
to have higher entropy.

The correlation of the  exponential
instability  with the various physical parameters of $N$-body systems 
suggests that the mixing properties  predicted by the existence of this 
instability may
play an important role in the evolution of gravitational systems.
Of particular importance is its effect on  
the time-scales of relaxation phenomena as originally pointed out
 by Gurzadyan
\& Savvidy~\cite{Gurzadyan and Savvidy1,Gurzadyan and Savvidy2}.

One has, however, to note the following.  The characteristic time-scales, 
which are found to be a fraction of a crossing time (Table~\ref{sphex:tabC}), 
need not be simply equal to the actual time scales of relaxation 
and physical evolution of a gravitational system.  
The latter properties are determined
by the rate of diffusion in the action variables which may be much slower 
than the g time \cite{Chirikov}.  However, since no
universal relaxation time scale does exist for all N-body systems and different 
physical quantities may have different relaxation time scales,
this does not imply that the exponential divergence has no physical observable effect. 
 Neither does it mean that the diffusion rate is determined for all action 
variables by classical relaxation theory since this assumes an
integrable system which remains so under perturbations due to 
discreteness --- and this contradicts the observed exponential divergence. 
Moreover, though the divergence is
local, it takes place {\em at almost all initial conditions} so that the 
structure of the phase space is such that,  for a given trajectory
most nearby trajectories diverge
{\it almost everywhere}. 
In this sense the divergence, although locally characterised,
is a global phenomenon and must have an effect on the statistical properties of the dynamical
system (see e.g. reference \cite{Chirikov}).
The criticism  by Goodman, 
Heggie and Hut \cite{Goodman et al.} of the linear theory of stability
(Jacobi equation) and its relation to macroscopic relaxation, 
is therefore removed due 
to these results from the theory of dynamical systems.
Cerruti-Sola and 
Pettini \cite{Cerruti-Sola and Pettini} suggest to add 
the first and second order time derivatives of the
potential energy in the Jacobi equation, which perhaps can be of some 
interest for some systems, but
not for stationary stellar systems where the  potential energy does not 
undergo significant time fluctuations.
Also, they state that due to Schur's theorem, if the curvatures are 
constant, one can replace 
the two-dimensional curvatures with the averaged (Ricci or scalar) curvatures 
without loss of information. 
This, however, does not follow from that theorem. 
Finally, their use of the Eisenhart metric is also not justified, since 
that metric is not positively defined.

These examples once more show the difficulty of the problem
of characterising and interpreting the instability of  nonlinear many dimensional system like
$N$-body gravitating systems. Among recent results in this area
one can also mention the indications for the possible existence of a
universal scaling in $N$-body dynamics, either of fractal 
nature \cite{Vega} or due to Feigenbaum period doubling 
bifurcations \cite{GMO}.

\section{Acknowledgements}
We thank Sverre Aarseth for  sending a copy of his NBODY2 code
and Simon Goodwin for helping with its use. 
During the period this work was 
undertaken V.G. was supported by the Royal Society and 
A.A.E.Z. by a British Foreign and Commonwealth Office Chevening Scholarship.
V.G. is thankful to John Barrow and the  late Roger Tayler for
hospitality at the University of Sussex Astronomy Centre.

\end{document}